# Magnetic Flux Transport Identification of Active Reconnection: MMS Observations in Earth's Magnetosphere

Yi Qi[1,2], Tak Chu Li[3], Christopher T. Russell[1], Robert E. Ergun[2], Ying-Dong Jia[1], and Mark Hubbert[1]
[1] Department of Earth, Planetary, and Space Sciences, University of California, Los Angeles, CA, USA; Yi.Qi@lasp.colorado.edu
[2] Laboratory for Atmospheric and Space Physics, University of Colorado Boulder, Boulder, CO, USA
[3] Department of Physics and Astronomy, Dartmouth College, Hanover, NH, USA

## Abstract

Magnetic reconnection plays an important role in converting energy while modifying field topology. This process takes place under varied plasma conditions during which the transport of magnetic flux is intrinsic. Identifying active magnetic reconnection sites with in situ observations is challenging. A new technique, Magnetic Flux Transport (MFT) analysis, has been developed recently and proven in numerical simulation for identifying active reconnection efficiently and accurately. In this study, we examine the MFT process in 37 previously reported electron diffusion region (EDR)/reconnection-line crossing events at the day-side magnetopause and in the magnetotail and turbulent magnetosheath using Magnetospheric Multiscale measurements. The coexisting inward and outward MFT flows at an X-point provides a signature that magnetic field lines become disconnected and reconnected. The application of MFT analysis to in-situ observations demonstrates that MFT can successfully identify active reconnection sites under complex varied conditions, including asymmetric and turbulent upstream conditions. It also provides a higher rate of identification than plasma outflow jets alone. MFT can be applied to in situ measurements from both single- and multi-spacecraft missions and laboratory experiments.

*Unified Astronomy Thesaurus concepts:* Solar magnetic reconnection (1504); Solar-terrestrial interactions (1473); Space plasmas (1544)

## 1. Introduction

Magnetic reconnection is a fundamental plasma process. During reconnection, the magnetic field lines change connectivity (Dungey 1961) and facilitate explosive energy conversion from magnetic to particle kinetic and thermal energies resulting in significant heating and acceleration (e.g., Treumann & Baumjohann 2013). Reconnection is ubiquitous in the universe. It occurs under various conditions, in both quiet and dynamic environments. Reconnection has been observed throughout the heliosphere, including the solar corona (Smartt et al. 1993; Xue et al. 2016) and solar wind (Gosling et al. 2005; Fargette et al. 2021; Phan et al. 2020). Near Earth, the magnetosphere provides a highly accessible natural laboratory for detailed measurements of reconnection events, which have been recently made by the Magnetospheric Multiscale (MMS) mission (Burch et al. 2016, Lavraud et al. 2016; Russell et al. 2017).

Reconnection can couple multiple scales (Hesse & Cassak 2020). The center of reconnection is the topological X-point/X-line, where the magnetic field both annihilates and reconnects. Around this site is a region on electron scales, known as the electron diffusion region (EDR), where both electrons and ions violate the frozen-in condition and are no longer dynamically coupled to the magnetic field (Ng et al. 2011; Bessho et al. 2014; Shay et al. 2016). Signatures of this region include deviation of the electron and ion bulk flow speed from the $E \times B$ drift, the strong current carried mainly by electrons, electron energization, enhanced energy conversion, crescent-shaped agyrotropic electron velocity distribution functions, and a small radius of curvature of the magnetic field (Büchner & Zelenyi 1991; Le et al. 2013; Webster et al. 2018; Tang et al. 2019). The larger region in which the EDR is embedded is the ion diffusion region (IDR), where ions are decoupled from the magnetic field while electrons remain coupled. Here ions are energized, and electrons form super-Alfvénic outflow jets. The separation between ions and electrons results in Hall electric and magnetic fields (Graham et al. 2016; Genestreti et al. 2020). At larger scales, Alfvénic ion outflow jets can be observed.

The identification of active reconnection has been challenging. With high-resolution observations from MMS, we are able to see most of the above-mentioned signatures. However, these signatures may not be observed coherently. The trajectory of the spacecraft has a significant influence on the observational profile. In addition, the asymmetry of upstream conditions (Shay et al. 2016), the existence of finite guide fields (Ng et al. 2011; Bessho et al. 2019), and the strong shear flows (Liu et al. 2018; Li et al. 2021) likely in a turbulent system may all distort the signatures, adding complexity to reconnection.

Recent studies have analyzed the transport of magnetic flux around an X-point in kinetic simulations (Liu & Hesse 2016; Liu et al. 2018). An innovative technique based on Magnetic Flux Transport (MFT) has been developed and carefully analyzed in a two-dimensional gyrokinetic simulation (Li et al. 2021), where MFT was applied to both symmetric and asymmetric reconnection X-point regions in turbulence generated by a double-vortex setup (Li et al. 2016). MFT successfully captures bidirectional inflows and outflows of magnetic flux around active X-points in a region significantly smaller than the region extended by plasma outflow jets or finite energy conversion, and thus the MFT method can locate the active reconnection sites more accurately than previous methods. In addition, although strong background shear flows distort the bidirectional plasma outflow jets, the velocity of MFT $U_\psi$ maintains its regular pattern, demonstrating that the MFT method is more robust than previous methods. Quadrupolar structures are observed in $\nabla \cdot U_\psi$ at the X-points, supporting the active reconnection picture. Based on these numerical modeling results, MFT has the potential to be a more accurate indicator of active reconnection. This study applies this newly developed technique to MMS in situ observations and validates its functionality under various plasma conditions.

Table 1
Event List of EDR/Reconnection-line Crossings

| Date and time | Location | Guide Field | Type | Spacecraft Separation [de] | Reference |
| --- | --- | --- | --- | --- | --- |
| **2015-10-16 13:07:02** | Day side | ∼0 | Classic | ∼6 | Burch et al. (2016) |
| **2015-12-08 11:20:43** | Day side | ∼1 | Classic | ∼6 | Burch & Phan (2016) |
| **2015-12-06 23:38:31** | Day side | ∼0.2 | Classic | ∼10 | Khotyaintsev et al. (2016) |
| **2015-10-25 11:07:46** | Sheath | ∼0.5 | Classic | ∼20 | Eriksson et al. (2018) |
| **2016-12-09 09:03:54** | Sheath | >5 | Electron-only | ∼5 | Phan et al. (2018) |
| **2016-11-09 13:39:26** | Shock transition region | ∼0 | Classic | ∼16 | Wang et al. (2019) |
| **2017-07-11 22:34:02** | Tail | ∼0 | Classic | ∼1 | Torbert et al. (2018) |
| **2017-06-17 20:24:07** | Tail | ∼0 | Electron-only | ∼4 | Lu et al. (2020) |
| **2017-08-10 12:18:33** | Tail | ∼0.1 | Classic | ∼2 | Zhou et al. (2019) |
| **31 EDRs** | Day side | varying | Classic | 2-70 | Webster et al. (2018) |

## 2. Data

The data used herein is obtained by the MMS mission. This mission is designed to capture the elusively thin and fast-moving diffusion regions of reconnection with unprecedented time resolution (Burch et al. 2015). The orbits cover the most common reconnection locations on both the day side and night side of Earth's magnetosphere, a natural laboratory for in situ observations of reconnection. Four identical spacecraft form a tetrahedron configuration with spacecraft separation varying from ∼10 to 150 km which provides spatial resolution down to kinetic scales (Fuselier et al. 2016). The magnetic field is measured by the fluxgate magnetometer (FGM) (Russell et al. 2016) at its highest sampling rate of 128 Hz.

We select previously identified EDR or reconnection-line crossing events to represent different environments of reconnection near Earth. The events include reconnection in the turbulent shock transition region (Wang et al. 2019), turbulent magnetosheath (Eriksson et al. 2018; Phan et al. 2018), dayside magnetopause (Burch et al. 2016; Burch & Phan 2016; Khotyaintsev et al. 2016), magnetotail (Torbert et al. 2018; Zhou et al. 2019; Lu et al. 2020), and a list of EDRs reported by Webster et al. (2018). The selected events are sufficiently typical, representing symmetric and asymmetric upstream conditions, varying guide field strength, quiet and turbulent regions, as well as classic ion-coupled reconnection and newly discovered electron-only reconnection. Requiring events with four spacecraft measurements for calculating spatial gradient, we exclude one event where MMS3 was not available from the 32 EDR events. We also note that three out of the nine single events considered overlap with the EDR list (Webster et al. 2018). Thus, in total there are 31 + (9 − 3) = 37 events that we apply the MFT analysis to, listed in Table 1, with the average spacecraft separation normalized to the electron inertial length $d_e$ based on upstream hybrid densities (Equation 17, Cassak & Shay 2007).

To apply this technique to observations by MMS and other spacecraft missions, we first validate it with known events. With the 37 events of varied plasma conditions, we find this technique robust enough for identifying reconnection.

## 3. Method

The MFT velocity $U_\psi$ was previously derived in one and two-dimensions (Liu & Hesse 2016; Liu et al. 2018). The formula of this velocity can be found in Li et al. (2021):

$$U_\psi \equiv v_{ep} - (v_{ep} \cdot \hat{b}_p)\hat{b}_p + \frac{cE'_{ez}}{B_p}(\hat{z} \times \hat{b}_p), \quad (1)$$

where $\hat{b}_p = B_p/B_p$ is the unit vector of the magnetic field component ($B_p$) in the 2D reconnection plane, the LN plane in LMN coordinates; $v_{ep}$ is the electron flow in the 2D reconnection plane; and $\hat{z}$ is the out-of-plane (M) direction. $E'_e = E + v_e \times B/c$ is the nonideal electric field in the electron frame. The first two terms represent the in-plane electron flow perpendicular to $B_p$, and the last term represents the slippage between magnetic flux and electron flow. Without separating the perpendicular electron flow and slippage terms, Equation (1) can be simplified to a form without electron velocity (Li et al. 2021):

$$U_\psi = (cE_z/B_p)(\hat{z} \times \hat{b}_p) \quad (2)$$

According to simulation work (Li et al. 2021), $U_\psi$ will form super-Alfvénic jets in both inflow (N) and outflow (L) directions, indicating strong MFT close to the active X-points. The upstream hybrid Alfvén speed $V_A$ is calculated using the L component of the magnetic field and ion density on the two upstream sides (Equation 13, Cassak & Shay 2007). We subtract the ion bulk flow velocity from $U_\psi$ (and electron flow) to demonstrate the MFT jets more clearly in the ion frame.

With four spacecraft measurements, we can estimate the divergence of the magnetic flux transport velocity $\nabla \cdot U_\psi$ following the linear gradient technique (Chanteur, ISSI, 1998,

Ch. 11). This quantity is able to represent the converging inflows and diverging outflows of magnetic flux. In addition to the signature in $U_\psi$, these bidirectional inflows and outflows of magnetic flux at an X-point signify active reconnection.

Transforming the field data to the LMN coordinate is required for the MFT analysis. For the nine single events, the coordinate rotation matrices from the literature are used. For events from Webster et al. (2018), the Minimum Variance Analysis (MVA) technique (Sonnerup & Cahill 1967) is applied to the magnetic field in the interval around the recorded EDR crossings to determine the LMN coordinate. The maximum variance direction (L) aligns with the reconnecting field direction, corresponding to the expected outflow direction. The minimum variance direction (N) gives the normal of the reconnecting current sheet, the expected inflow direction.

There are two signatures of active reconnection in MFT analysis. They are (1) coexisting magnetic flux ($U_\psi$) jets in the inflow (N) and outflow (L) directions, and (2) a significantly enhanced divergence of flux transport ($\nabla \cdot U_\psi$) at the X-point. Previous theory and simulation work suggested that $U_\psi$ is at least ion Alfvénic ($\gtrsim 0.5\, V_A$) and the divergence on the order of 0.1 electron cyclotron frequency ($\nabla \cdot U_\psi \gtrsim 0.05\, f_{ce}$) (Li et al. 2021). Observing either signature identifies an encounter of an active reconnection site. For each event, we select a region of interest around the EDR/reconnection-line crossing such that the radius of the field line curvature $R_c$ becomes $\lesssim$ the ion gyroradius $\rho_i$, indicating agyrotropic ion motions indicative of the IDR (Rogers et al. 2019). Within the selected region, we record the peak values of $U_\psi$ jets and $\nabla \cdot U_\psi$.

## 4. Results

### 4.1. Example of a Successfully Identified Reconnection Site

As an example of MFT signatures in MMS data, we summarize in Figure 1 the analysis of an active reconnection event in the magnetosheath reported by Eriksson et al. (2018). At the reversal of the magnetic field L component (a), the magnetic curvature increases and $R_c$ reaches the electron gyroradius $\rho_e$. The MFT velocity demonstrates a bipolar signature in the N direction (blue, (f)) and a unipolar peak in L (red). Both peaks exceed the upstream Alfvén speed $V_A$ (horizontal dotted line). These bidirectional MFT inflows and unidirectional outflow are consistent with (h) the deduced spacecraft trajectory of MMS crossing the actively reconnecting current sheet from upstream to downstream and then upstream on the other side. In panel (g) the bipolar structure in $\nabla \cdot U_\psi$ is consistent with converging MFT inflows and diverging outflows near the X-point, with a peak value exceeding the order of $0.1\, f_{ce}$. The observed MFT signatures agree well with simulation.

We plot the L and N components of $U_\psi$ on each spacecraft in Figure 2. The patterns of the velocity are similar on all spacecraft, suggesting the scale of the structure is greater than the spacecraft separation. Thus, MMS resolves the structure of $U_\psi$. MMS4 and MMS2 detect the two strongest peaks in the L component, in agreement with the X-point being south of the spacecraft and MMS4 and MMS2 being the closest to it. Examining $U_\psi$ on each spacecraft is generally useful for events with a large spacecraft separation, where the four-spacecraft average may not resolve the structure of $U_\psi$.

### 4.2. Identification in 37 Events

The same analysis is applied to the 37 events. Figure 3 summarizes the result. Events are ordered from left to right with increasing normalized spacecraft separation.

One of the two MFT signatures for active reconnection is coexisting Alfvénic jets in the inflow and outflow directions. Figure 3(a) shows the peaks of MFT jets in the N (circles) and L (crosses) directions for all events. $U_\psi$ is normalized to $V_A$. The dashed lines mark $\pm 0.5\, V_A$, above which they are considered Alfvénic. Jets less than $0.5\, V_A$ are masked. For almost all events, MMS observed coexisting N and L jets, demonstrating both inflow and outflow MFT in these events. Only two events on 2016 November 18 and February 7 do not have this signature. For the event on 2016 February 7, examining $U_\psi$ on each spacecraft also reveals coexisting super-Alfvénic inflow and outflow jets. For the event on 2016 November 18, the divergence of $U_\psi$ can be used and also identifies active reconnection (Figure 3(b)).

The other MFT signature of active reconnection is the divergence of $U_\psi$ being of the order of 0.1 of the local electron cyclotron frequency $f_{ce}$ or higher. Plotted in Figure 3(b) is $\nabla \cdot U_\psi$ normalized to $f_{ce}$. The dashed lines indicate $\pm 0.05\, f_{ce}$. Values smaller than 0.05 are considered below order $0.1\, f_{ce}$ and masked. Similar to the MFT jets, in nearly all events, the peak $\nabla \cdot U_\psi$ exceeds the threshold of $O(0.1\, f_{ce})$, confirming active reconnection encounters. The typical $\nabla \cdot U_\psi$ signature lies within $\pm 0.5\, f_{ce}$, consistent with the expected ordering. As discussed, the event on 2016 November 18 has a high enough divergence of $U_\psi$ as a signature of active reconnection. In total, all 37 events are successfully identified as active reconnection through MFT signatures.

The median value of the $\pm L$ jet peak is $3.1/-2.3\, V_A$ and that of the N jet peak is $2.0/-2.5\, V_A$. They indicate that typical MFT jets are super-Alfvénic around the active reconnection site. The median absolute value of $\nabla \cdot U_\psi$ is $0.3\, f_{ce}$, also meeting the criteria of being of order $0.1\, \Omega_e$ or higher.

## 5. Discussion

The two MFT properties that serve as active reconnection signatures are coexisting Alfvénic inflow and outflow flux jets and a high divergence of flux transport velocity. These two properties are observed in 97% and 95% of all the events. Satisfying either one of the two criteria will be sufficient for reconnection identification. In this case, all 37 events are successfully identified by MFT properties.

The MFT signatures are also compared with plasma outflow jet signatures. To confidently identify reconnection-line crossing using plasma flows, we usually need the flow to be bidirectional. The occurrence rates for Alfvénic bidirectional plasma outflow jets are lower than 50% (19% for ion and 43% for electron). If we use bidirectional outflow jets as the criteria for active reconnection crossing, we will very likely miss more than half of the events. Using only one jet as the criteria, it is not confined as closely around the X-line as the MFT signatures (Li et al. 2021), and it is harder to establish a straightforward link between the observation of a single plasma jet and the reconnection.

The existence of a finite guide field will make the situation more complicated. It does not only modify the topology at the reconnection sites but also possibly change the planar picture into a more turbulent three-dimensional scenario (Ng et al.

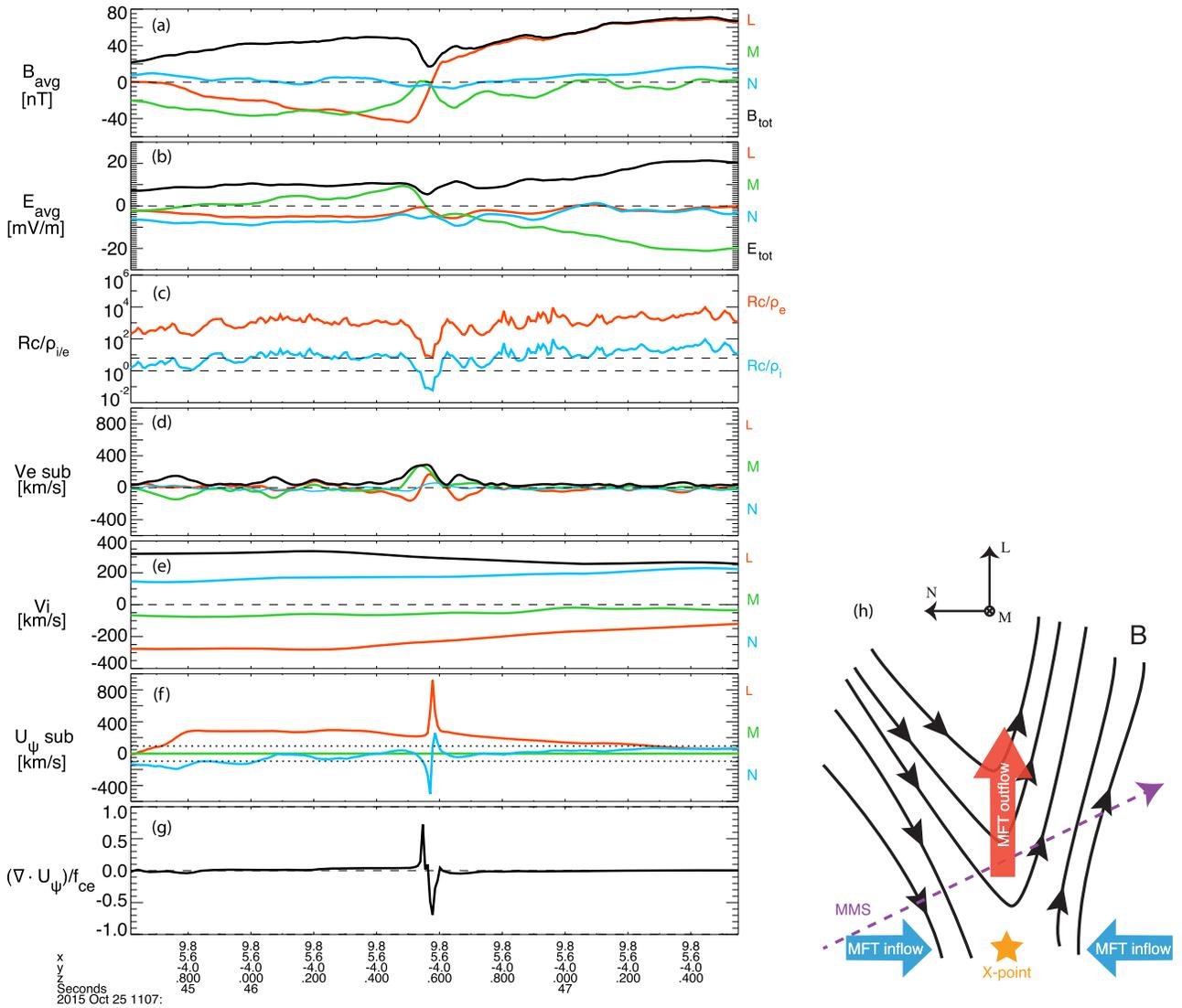

**Figure 1.** MMS observations of an example event on 2015 October 25. Vectors are transformed in LMN coordinates ($L = [0.31, -0.91, 0.28]$, $M = [0.31, 0.37, 0.87]$, $N = [-0.90, -0.19, 0.40]$ in GSE). Four-spacecraft-averaged (a) magnetic field and (b) electric field. (c) Radius of curvature $R_c$ normalized to the electron (red) and ion (blue) gyroradius. (d), (e) Electron bulk flow velocity (with ion velocity subtracted) and ion bulk flow velocity. (f) MFT velocity $U_\psi$ (with ion velocity subtracted). (g) $\nabla \cdot U_\psi$ normalized to the local electron cyclotron frequency $f_{ce}$ (h) Sketch of the MMS trajectory and expected MFT flows, adapted from Eriksson et al. (2018).

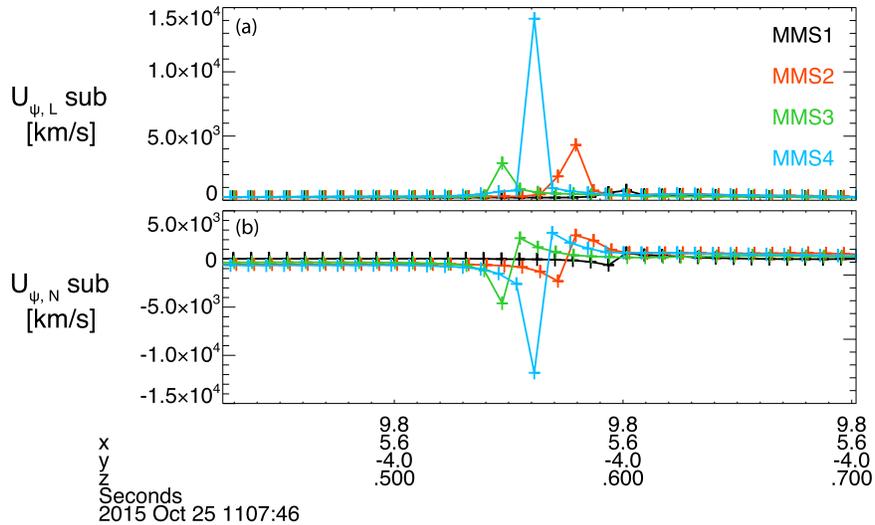

**Figure 2.** MFT velocity on each spacecraft. (a) The L and (b) N components of $U_\psi$ measured by the four spacecraft.

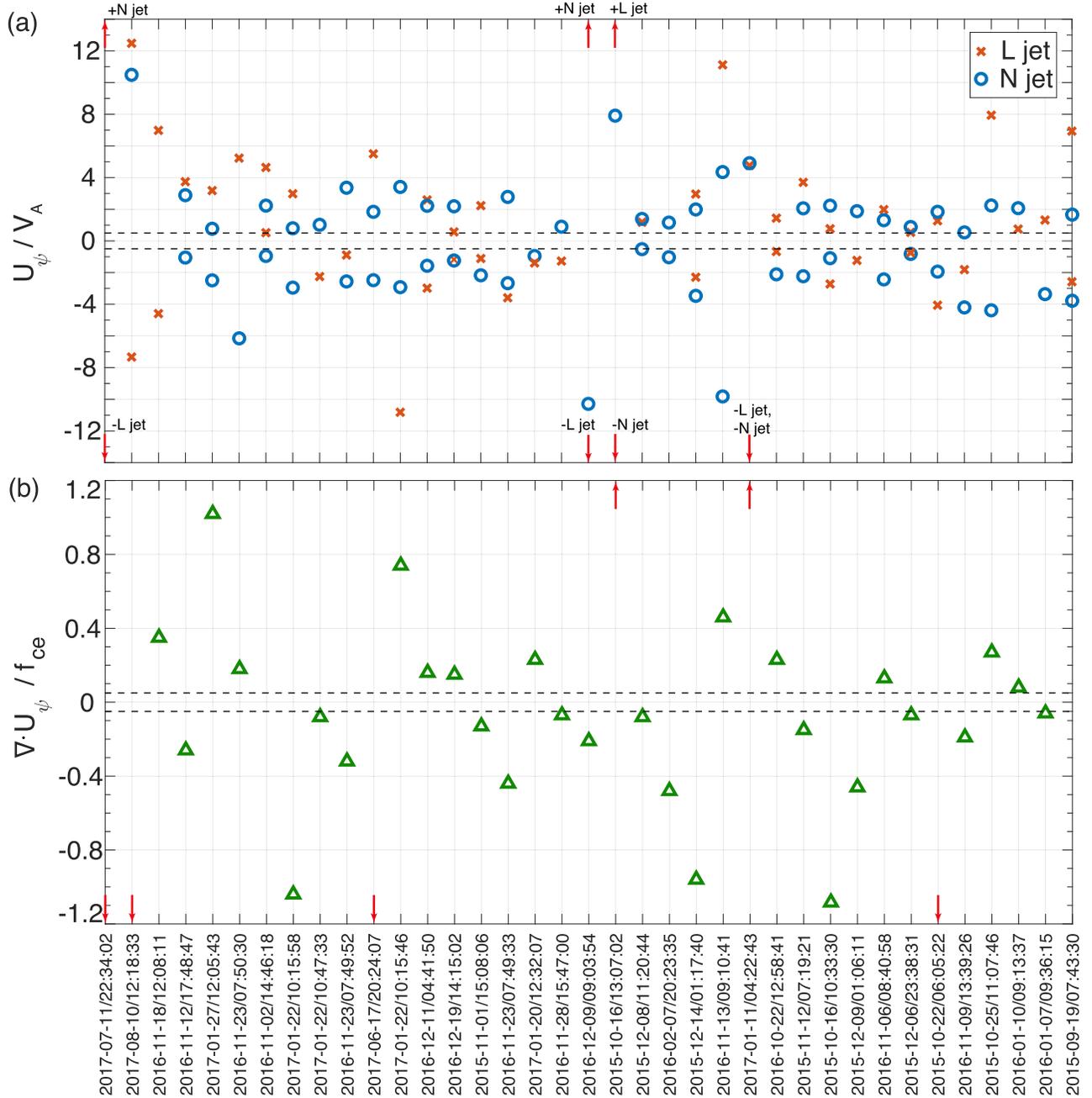

**Figure 3.** MFT signatures in 37 events. The peak values of (a) $U_\psi$ in the L and N directions normalized to $V_A$ and (b) $\nabla \cdot U_\psi$ normalized to $f_{ce}$. The red arrows indicate L/N jets and $\nabla \cdot U_\psi$ out of the plotted range. The dashed lines represent (a) $\pm 0.5\, V_A$ and (b) $\pm 0.05\, f_{ce}$.

2011; Chen et al. 2020). Our database includes a variety of guide field strengths, and as mentioned earlier, regardless of the guide field strength, the MFT analysis successfully identified active reconnection. Thus, we suggest the accuracy of MFT is robust in the presence of varying guide fields.

In classic reconnection, the IDR is typically elongated in the outflow direction. In a steady state, the large aspect ratio of this region translates into much higher outflow than inflow jets (by continuity of density and incompressibility). However, MFT flows tend to be highly localized to the X-line. This results in a smaller aspect ratio of the MFT flow region and therefore similar amplitudes for MFT inflows and outflows. The result in Figure 3(a), where the L and N jets are of the same order of magnitude, is consistent with this picture and also consistent with simulation.

Regardless of the broad range of normalized spacecraft separations of the 37 events (Table 1), $U_\psi$ ranges from the order of ion Alfvén to electron Alfvén speeds, and $\nabla \cdot U_\psi$ is of order 0.1–1 $f_{ce}$ (Figure 3), consistent with simulation (Li et al. 2021).

The choice of LMN coordinates has an impact on MFT analysis. In addition to using MVA on the magnetic field, MVA can also be applied to the electron velocity. Other methods including Minimum Faraday Residue (MFR) (Khrabrov & Sonnerup 1998; Haaland et al. 2019) and Maximum Directional Derivative (MDD) (Shi et al. 2019) may be considered. These methods have been compared in detail in recent studies (Denton et al. 2018; Genestreti et al. 2018). While determining the suitable LMN coordinates for MFT is out of the scope of this study, it will be investigated in the future.

## 6. Conclusions

In this study, for the first time we have applied the newly developed MFT technique to MMS data and showed this technique can successfully identify active reconnection. The two MFT properties that are signatures of active reconnection encounter are coexisting Alfvénic inflow and outflow flux jets and a high divergence of the flux transport velocity. The detection of either one is sufficient for identification. We select 37 previously reported EDR/reconnection-line crossing events on Earth's day-side magnetopause, in the magnetotail and turbulent magnetosheath, to test the capability of MFT under various plasma conditions. All events are successfully identified with the two MFT properties. The median value of the magnetic flux transport velocity is typically super-Alfvénic, sufficient for locating the active reconnection region. The divergence of the flux transport velocity has a median absolute value of $0.27 f_{ce}$, above the expected threshold for reconnection. The occurrence rates of these two properties are 97% and 95%, much higher success rates compared to using bidirectional plasma outflow jets. This application of MFT to the terrestrial data demonstrated its capability to identify reconnection under complex varied plasma conditions, motivating the application of this technique for analyzing reconnection in more heliospheric contexts such as the solar corona and solar wind turbulence, and in laboratory experiments.

We appreciate the efforts of the entire MMS team, including many individuals in spacecraft operation and data acquisition. We thank T. Phan, S. Schwartz, and S. Eriksson for helpful discussions. This research was supported by NASA MMS mission NNG04EB99C and NSF award AGS-2000222. We also acknowledge support from NASA grant 80NSSC21K1691. The work at UCLA was supported through subcontract 06-001 with the University of New Hampshire.

## ORCID iDs

Yi Qi ⓘ https://orcid.org/0000-0002-0959-3450
Tak Chu Li ⓘ https://orcid.org/0000-0002-6367-1886
Christopher T. Russell ⓘ https://orcid.org/0000-0003-1639-8298
Ying-Dong Jia ⓘ https://orcid.org/0000-0002-1631-291X

## References

Bessho, N., Chen, L.-J., & Shuster, J. R. 2014, GeoRL, 41, 8688
Bessho, N., Chen, L.-J., Wang, S., & Hesse, M. 2019, PhPl, 26, 082310
Büchner, J., & Zelenyi, L. M. 1991, AdSpR, 11, 177
Burch, J. L., Moore, T. E., Torbert, R. B., & Giles, B. L. 2015, SSRv, 199, 5
Burch, J. L., & Phan, T. D. 2016, GeoRL, 43, 8327
Burch, J. L., Torbert, R. B., Phan, T. D., et al. 2016, Sci, 352, aaf2939
Cassak, P. A., & Shay, M. A. 2007, PhPl, 14, 102114
Chen, L.-J., Wang, S., Contel, O. L., et al. 2020, PhRvL, 125, 025103
Denton, R. E., Sonnerup, B. U. Ö., Russell, C. T., et al. 2018, JGRA, 123, 2274
Dungey, J. W. 1961, PhRvL, 6, 47
Eriksson, E., Vaivads, A., Graham, D. B., et al. 2018, GeoRL, 45, 8081
Fargette, N., Lavraud, B., Rouillard, A., et al. 2021, A&A, 650, A11
Fuselier, S. A., Lewis, W. S., Schiff, C., et al. 2016, SSRv, 199, 77
Genestreti, K. J., Liu, Y.-H., Phan, T.-D., et al. 2020, JGRA, 125, e27985
Genestreti, K. J., Nakamura, T. K. M., Nakamura, R., et al. 2018, JGRA, 123, 9130
Gosling, J. T., Skoug, R. M., McComas, D. J., et al. 2005, JGR, 110, A01107
Graham, D. B., Khotyaintsev, Yu. V., Norgren, C., et al. 2016, GeoRL, 43, 4691
Haaland, S., Runov, A., Artemyev, A., & Angelopoulos, V. 2019, JGRA, 124, 3421
Hesse, M., & Cassak, P. A. 2020, GeoRL, 125, e25935
Khotyaintsev, Yu. V., Graham, D. B., Norgren, C., et al. 2016, GeoRL, 43, 5571
Khrabrov, A. V., & Sonnerup, B. U. Ö. 1998, GeoRL, 25, 2373
Lavraud, B., Zhang, Y. C., Vernisse, Y., et al. 2016, GeoRL, 43, 3042
Le, A., Egedal, J., Ohia, O., et al. 2013, PhRvL, 110, 135004
Li, T. C., Howes, G. G., Klein, K. G., et al. 2016, ApJL, 832, L24
Li, T. C., Liu, Y.-H., & Qi, Y. 2021, ApJL, 909, L28
Liu, Y.-H., & Hesse, M. 2016, PhPl, 23, 060704
Liu, Y.-H., Hesse, M., Guo, F., Li, H., & Nakamura, T. K. M. 2018, PhPl, 25, 080701
Lu, S., Wang, R., Lu, Q., et al. 2020, NatCo, 11, 5049
Ng, J., Egedal, J., Le, A., Daughton, W., & Chen, L.-J. 2011, PhRvL, 106, 065002
Phan, T. D., Bale, S. D., Eastwood, J. P., et al. 2020, ApJS, 246, 34
Phan, T. D., Eastwood, J. P., Shay, M. A., et al. 2018, Natur, 557, 202
Rogers, A. J., Farrugia, C. J., & Torbert, R. B. 2019, JGRA, 124, 6487
Russell, C. T., Anderson, B. J., Baumjohann, W., et al. 2016, SSRv, 199, 189
Russell, C. T., Strangeway, R. J., Zhao, C., et al. 2017, Sci, 356, 960
Shay, M. A., Phan, T. D., Haggerty, C. C., et al. 2016, GeoRL, 43, 4145
Shi, Q. Q., Tian, A. M., Bai, S. C., et al. 2019, SSRv, 215, 35
Smartt, R. N., Zhang, Z., & Smutko, M. F. 1993, SoPh, 148, 139
Sonnerup, B. U. Ö., & Cahill, L. J. 1967, JGR, 72, 171
Tang, B.-B., Li, W. Y., Graham, D. B., et al. 2019, GeoRL, 46, 3024
Torbert, R. B., Burch, J. L., Phan, T. D., et al. 2018, Sci, 362, 1391
Treumann, R. A., & Baumjohann, W. 2013, FrP, 1, 31
Wang, S., Chen, L., Bessho, N., et al. 2019, GeoRL, 46, 562
Webster, J. M., Burch, J. L., Reiff, P. H., et al. 2018, JGRA, 123, 4858
Xue, Z., Yan, X., Cheng, X., et al. 2016, NatCo, 7, 11837
Zhou, M., Deng, X. H., Zhong, Z. H., et al. 2019, ApJ, 870, 34